\documentclass[epj]{svjour}
\usepackage{graphics}
\usepackage{xcolor}
\usepackage[utf8]{inputenc}
\usepackage{amsmath}
\usepackage{amsfonts}
\usepackage{cite} 
\usepackage{breqn}
\usepackage{graphicx}
\usepackage{hyperref}
\allowdisplaybreaks

\providecommand{\Rie}[3]{\mathcal{R}_{#1}{}^{#2}{}_{#3}}
\providecommand{\ctG}[3]{\Gamma_{#1}{}^{#2}{}_{#3}}
\providecommand{\ctg}[3]{\gamma_{#1}{}^{#2}{}_{#3}}
\providecommand{\B}[3]{\mathcal{B}_{#1}{}^{#2}{}_{#3}}

\providecommand{\A}[1]{\mathcal{A}_{#1}}
\providecommand{\Ri}[1]{\mathcal{R}_{#1}}
\providecommand{\deV}[1]{\mathrm{d}V^{#1}}

\begin{document}
\title{Cosmological Solutions in Polynomial Affine Gravity with Torsion}
%\subtitle{Do you have a subtitle?\\ If so, write it here}
\author{Oscar Castillo-Felisola \inst{1,2} 
        \and Bastian Grez \inst{1} 
        \and Gonzalo J. Olmo \inst{3} 
        \and Oscar Orellana \inst{4} 
        \and Jos\'e Perdiguero G\'arate \inst{3}
}                    % Do not remove
%
%\offprints{}          % Insert a name or remove this line
%
\institute{Departamento de F\'isica, Universidad T\'ecnica Federico Santa Mar\'ia Casilla 110-V, Valpara\'iso, Chile \and
Centro Cient\'ifico Tecnol\'ogico de Valpara\'iso Casilla 110-V, Valpara\'iso, Chile \and
Departamento de F\'isica Te\'orica and IFIC, Centro Mixto Universidad de
Valencia - CSIC. Universidad de Valencia, Burjassot-46100, Valencia, Spain \and 
%Departamento de Astronom\'ia y Astrof\'isica, Universitat de Valencia, Dr. Moliner 50, 46100, Burjassot (Valencia), Spain \and
Departamento de Matem\'aticas, Universidad T\'ecnica Federico Santa Mar\'ia Casilla 110-V, Valpara\'iso, Chile}
\date{Received: date / Revised version: date}
% The correct dates will be entered by Springer
%
\abstract{The Polynomial Affine Gravity is an alternative gravitational model, where the interactions are mediated solely by the affine connection, instead of the metric tensor. In this paper, we explore the space of solutions to the field equations when the torsion fields are turned on, in a homogeneous and isotropic (cosmological) scenario. We explore various metric structures that emerge in the space of solutions.
 % end of PACS codes
} %end of abstract
\maketitle

\section{Introduction}
\label{sec:Introduction}

Einstein's theory of General Relativity (GR) is currently the most successful theory to describe gravitational interactions, exhibiting excellent agreement between theoretical predictions and observational data in a variety of scenarios \cite{Will_2014,will_2018,Weinberg:2008zzc}, from laboratory and solar system scales, to the orbital motions of binary pulsars, cosmology and even gravitational waves from colliding compact objects \cite{Abbott_2016,Abbott_2017} and the shadows of supermassive black holes. 

In its traditional formulation, GR is a theory in which the gravitational interaction is solely mediated by the 
metric tensor \cite{Einstein_GR,Einstein_GR_Bases}, from which other quantities such as a covariant derivative or curvature tensors can be derived. The nonlinear character of the equations make it necessary to use certain strategies and impose symmetries to obtain a simplified form that allows us to perform explicit computations. In cosmological scenarios, for instance, homogeneity and isotropy \cite{Frie,Friedmann,Lema,Lema_Expansion,Robertson_1,Robertson_2,Robertson_3} turn the original set of coupled, nonlinear differential equations for ten independent variables into a second-order equation for a single function, the scale factor  \cite{Einstein_GR_Feqs}. The evolution of this factor is determined by the matter-energy sources, which may include a cosmological constant \cite{Einstein_Cosmological_Constant} or some other form of dark energy, and the curvature of the spatial sections of the foliated space-time. 

Despite its success, GR also faces difficulties that suggest that a more fundamental description of the gravitational interactions is necessary. Its combination with quantum theory indicates that an ultraviolet completion is needed \cite{PhysRev.160.1113,PhysRev.162.1195,PhysRevD.10.401,PhysRevD.10.411}, while the need to include a dark sector, for which no direct evidence exists, that dominates the cosmic evolution and the dynamics of galaxies and clusters \cite{Rotation_Curve,Rotation_Curve_2,NASERI2021100888,10.1093/mnras/stz2757,Riess_1998}, may point towards new gravitational dynamics in the infrared.  As a result, it is becoming generally accepted that GR should be viewed as an effective theory that may require extensions at very short and very large length scales. Obviously, the kind of modifications needed are the million dollar question, and multiple alternatives can be found in the literature. Among those, theories of the $f(R)$ type \cite{10.1093/mnras/150.1.1,STAROBINSKY198099,Hehl_1995,Baldazzi_2022,Vitagliano_2011,Karahan_2012,SARDANASHVILY_2011,OLMO_2011}, scalar-tensor theories, and various extensions of them \cite{ASENS_1924_3_41__1_0,ASENS_1925_3_42__17_0,Kaluza_Klein,Klein,saridakis2023modified,Shankaranarayanan_2022}, among others, have become very popular in the last two decades for theoretical and phenomenological reasons, offering a variety of strategies and alternative mechanisms to justify the accelerated cosmic expansion, inflationary scenarios, and the possible existence of exotic compact objects. 

A fundamental ingredient in the construction of any gravity theory is the type of fields associated with the gravitational interaction. The traditional approach assumes that the underlying geometry is Riemannian (or pseudo-Riemannian), being described solely by the metric tensor. Alternative approaches in which metric and connection are treated as equally fundamental and independent fields are also gaining attention in the last years \cite{Olmo:2011uz}. The role that torsion and non-metricity could have at cosmic scales and in strong gravity scenarios offers a window to explore new gravitational phenomena beyond the Riemannian framework \cite{Bahamonde:2015zma, Bahamonde:2022kwg,Bahamonde:2021srr,BeltranJimenez:2017tkd,Harko:2018gxr,Silva:2022pfd}. This can be viewed as a complementary approach to the modified theories scenario but considering modified geometry instead. 

The freedom contained in the connection, with up to 64 independent components, offers a vast range of options to generate new gravitational phenomena and even to potentially accommodate adaptations needed to build a framework more suitable to incorporate quantum phenomena \cite{Hehl:1994ue}. In this sense, it is important to note that the most successful description of the fundamental interactions is based on gauge theories, which describe the dynamics of the connections of the symmetry groups of the standard model of elementary particles. It is thus natural to question if a purely connection-based formulation of gravity is possible, such that it could be represented in a form more closely related to the other interactions. The solution to this question is far from trivial, though some examples of purely affine theories exist in the literature.  

Looking back at the literature on purely affine gravity theories, the model proposed by  Sir Arthur  Eddington and Einstein \cite{Eddington1923-EDDTMT,schrodinger1985space,Einstein:1923:AFG} represents a simple example that nonetheless illustrates the key challenges faced by this type of theories. Eddington's theory is defined by the square root of the determinant of the symmetric part of the Ricci tensor, which is a diffeomorphism invariant quantity \cite{eisenhart1972non}. The variation of this action (with respect to the affine connection) can be manipulated to obtain the well-known Einstein equations in vacuum coupled to an arbitrary cosmological constant (see, for instance, \cite{POP_AWSKI_2007}, where the role of the antisymmetric part of the Ricci is also analyzed). As a result, the Ricci tensor can be interpreted as an emergent metric tensor (as long as the cosmological constant does not vanish). Despite its formal resemblance with GR, Eddington's theory faces evident difficulties when trying to couple gravity to matter fields, as there is no clear mechanism to build a suitable matter action in the absence of a metric tensor. Some recent attempts in this direction can be found in \cite{Knorr_2021,POP_AWSKI_2008,Pop_awski_2009,Filippov_2010,Azri_2015}, where  a metric tensor, that couples only to the matter sector is considered. Other formulations of purely affine theories have been inspired by the canonical approach to quantum gravity \cite{Krasnov_2011}, where the only dynamical field is an $SU(2)$ connection. Eddington's theory has also inspired metric-affine formulations  in recent years in the form of determinantal Born-Infeld like actions \cite{BI_Gravity,Deser_1998,Vollick:2003qp,Banados:2010ix,Jim_nez_2021,Afonso:2021aho}. 

An alternative approach to the purely affine formulation of gravity is provided by the Polynomial Affine Gravity (PAG) model. The PAG action is designed following a reasoning that parallels the \textit{dimensional analysis} technique of field theories, considering the irreducible terms that can be constructed out of the affine connection and its first derivatives and that preserve the invariance under diffeomorphisms. Because of the absence of a metric tensor, there is a geometric constraint to satisfy in order to build the most general scalar densities in the affine geometry, which leads to a finite number of terms in the action. This property is usually referred to as the \textit{rigidity} of the model.

Moreover, it is possible to couple a scalar field to the affine model using the same \textit{dimensional analysis} principle and, as a consequence, the \textit{rigidity} of the model is inherited by the coupling mechanism. This approach avoids the use of a metric to build the matter action, bypassing also the difficulties of Eddington's approach, and provides a new landscape to build purely affine gravity theories. The model has been studied in Refs.~\cite{castillofelisola2016polynomial,castillofelisola2016einsteins,castillofelisola2019cosmological,Castillo_Felisola_2018,Castillo_Felisola_2020,Castillo_Felisola_2022_EPJC,Castillo_Felisola_2022_Universe} and in this work, we explore the consequences of including torsion effects coming from the antisymmetric part of the affine connection in a four-dimensional cosmological setting.

The paper is organized as follows: In Section \ref{sec:PAG}, we present a brief overview of the model, highlighting its features and how to build up the ansatz for the affine connection. In Section \ref{sec:solutions} we present a complete scan of cosmological solutions to the field equations. In Section \ref{sec:analysis}, we analyses and discuss the cosmological solutions, and their physical interpretation by obtaining descendant (or emerging) metric structures coming from the irreducible components of the affine connection. Final remarks are presented in Section \ref{sec:final_remarks}. 

\section{Polynomial Affine Gravity}
\label{sec:PAG}

As said earlier, the Polynomial Affine Gravity is an alternative gravitational model whose fundamental field is  the affine connection, endowing the manifold only with an affine structure $(\mathcal{M},\Gamma)$. In order to build the action, it is convenient to decomposes the affine connection as
\begin{equation}
\begin{aligned}
    \label{affine_connection}
    \hat{\Gamma}_{\alpha}{}^{\beta}{}_{\gamma} & = \hat{\Gamma}_{(\alpha}{}^{\beta}{}_{\gamma)} +  \hat{\Gamma}_{[\alpha}{}^{\beta}{}_{\gamma]},  \\
    & = \ctG{\alpha}{\beta}{\gamma} + \B{\alpha}{\beta}{\gamma} + \delta^{\beta}_{[\gamma}\A{\alpha]},
\end{aligned}
\end{equation}
where the first term corresponds to the symmetric part of the connection $\hat{\Gamma}_{(\alpha}{}^{\beta}{}_{\gamma)} = \ctG{\alpha}{\beta}{\gamma}$, and
the last two terms are related to the torsion tensor. The former represents  the purely tensorial (traceless) part of the torsion $\B{\alpha}{\beta}{\gamma}$, while the latter is a pure vectorial object $\A{\alpha}$.
Additionally,  the introduction of the volume element is necessary and, in the absence of a metric tensor, one can use the wedge product to define it as  $\deV{\alpha\beta\gamma\delta} = \mathrm{d}x^{\alpha}\wedge\mathrm{d}x^{\beta}\wedge
\mathrm{d}x^{\gamma}\wedge\mathrm{d}x^{\delta}$. It is worth emphasizing that the action must preserve the invariance under diffeomorphisms, which is broken by the symmetric part of the affine connection, and as a consequence of this,
the symmetric part must appear in the action only through the covariant derivative $\ctG{\alpha}{\beta}{\gamma} \to \nabla^\Gamma$.
Therefore, the fundamental building blocks of the affine model are 
\begin{equation}
\label{irreducible_fields}
\nabla^\Gamma_\alpha, \B{\alpha}{\beta}{\gamma}, \mathcal{A}_\alpha, \mathrm{d}V^{\alpha\beta\gamma\delta}.
\end{equation}

In order to build up the action, we use a sort of \textit{dimensional analysis} technique which has been reviewed in \cite{castillofelisola2016einsteins,Castillo_Felisola_2020}.
The method allows one to consider every scalar densities composed by powers of Eq. \eqref{irreducible_fields} by using the operators $\mathcal{N}$ and $\mathcal{W}$ which count the number of free indices and the weight of the field, respectively. The analysis provides a geometrical constraint equation that limits the number of configurations. For each configuration, there are multiple permutations that need to be analyzed by taking into account the symmetries of the fundamental fields (see \cite{castillofelisola2016einsteins,Castillo_Felisola_2020}  for more details on this procedure). 
A three-dimensional version of this model was developed following this approach in Refs. \cite{Castillo_Felisola_2022_EPJC,Castillo_Felisola_2022_Universe}.

The most general action (up to topological invariants and boundary terms) in four dimensions is given by
\begin{equation}
\label{PAG_action}
\begin{split}
S & = \int  \mathrm{d}V^{\alpha \beta \gamma \delta} \bigg[
      B_1 \mathcal{R}_{\mu\nu}{}^{\mu}{}_{\rho}\mathcal{B}_{\alpha}{}^{\nu}{}_{\beta}\mathcal{B}_{\gamma}{}^{\rho}{}_{\delta}
    + B_2 \mathcal{R}_{\alpha\beta}{}^{\mu}{}_{\rho} \mathcal{B}_{\gamma}{}^{\nu}{}_{\delta} \mathcal{B}_{\mu}{}^{\rho}{}_{\nu}
    \\
    & \quad
    + B_3 \mathcal{R}_{\mu\nu}{}^{\mu}{}_{\alpha} \mathcal{B}_{\beta}{}^{\nu}{}_{\gamma} \mathcal{A}_\delta
    + B_4 \mathcal{R}_{\alpha\beta}{}^{\sigma}{}_{\rho}\mathcal{B}_{\gamma}{}^{\rho}{}_{\delta}\mathcal{A}_\sigma
    \\
    & \quad
    + B_5 \mathcal{R}_{\alpha \beta}{}^{\rho}{}_{\rho} \mathcal{B}_{\gamma}{}^{\sigma}{}_{\delta} \mathcal{A}_\sigma
    + C_1 \mathcal{R}_{\mu\alpha}{}^{\mu}{}_{\nu} \nabla_\beta \mathcal{B}_{\gamma}{}^{\nu}{}_{\delta}
    \\
    & \quad
    + C_2 \mathcal{R}_{\alpha\beta}{}^{\rho}{}_{\rho} \nabla_\sigma \mathcal{B}_{\gamma}{}^{\sigma}{}_{\delta}
    + D_1 \mathcal{B}_{\nu}{}^{\mu}{}_{\lambda} \mathcal{B}_{\mu}{}^{\nu}{}_{\alpha} \nabla_\beta \mathcal{B}_{\gamma}{}^{\lambda}{}_{\delta}
    \\
    & \quad
    + D_2 \mathcal{B}_{\alpha}{}^{\mu}{}_{\beta} \mathcal{B}_{\mu}{}^{\lambda}{}_{\nu} \nabla_{\lambda} \mathcal{B}_{\gamma}{}^{\nu}{}_{\delta}
    + D_3 \mathcal{B}_{\alpha}{}^{\mu}{}_{\nu}\mathcal{B}_{\beta}{}^{\lambda}{}_{\gamma} \nabla_\lambda \mathcal{B}_{\mu}{}^{\nu}{}_{\delta}
    \\
    & \quad
    + D_4 \mathcal{B}_{\alpha}{}^{\lambda}{}_{\beta}\mathcal{B}_{\gamma}{}^{\sigma}{}_{\delta}\nabla_\lambda \mathcal{A}_\sigma
    + D_5 \mathcal{B}_{\alpha}{}^{\lambda}{}_{\beta} \mathcal{A}_\sigma \nabla_\lambda \mathcal{B}_{\gamma}{}^{\sigma}{}_{\delta}
    \\
    &\quad
    + D_6 \mathcal{B}_{\alpha}{}^{\lambda}{}_{\beta}\mathcal{A}_\gamma \nabla_\lambda A_\delta
    + D_7\mathcal{B}_{\alpha}{}^{\lambda}{}_{\beta} \mathcal{A}_\lambda \nabla_\gamma A_\delta
    \\
    & \quad
    + E_1\nabla_\rho \mathcal{B}_{\alpha}{}^{\rho}{}_{\beta} \nabla_\sigma \mathcal{B}_{\gamma}{}^{\sigma}{}_{\delta}
    + E_2 \nabla_\rho \mathcal{B}_{\alpha}{}^{\rho}{}_{\beta} \nabla_\gamma \mathcal{A}_\delta
    \\
    & \quad
    + F_1 \mathcal{B}_{\alpha}{}^{\mu}{}_{\beta} \mathcal{B}_{\gamma}{}^{\sigma}{}_{\delta} \mathcal{B}_{\mu}{}^{\lambda}{}_{\rho} \mathcal{B}_{\sigma}{}^{\rho}{}_{\lambda}
    + F_2\mathcal{B}_{\alpha}{}^{\mu}{}_{\beta} \mathcal{B}_{\gamma}{}^{\nu}{}_{\lambda} \mathcal{B}_{\delta}{}^{\lambda}{}_{\rho} \mathcal{B}_{\mu}{}^{\rho}{}_{\nu}
    \\
    &\quad
    + F_3 \mathcal{B}_{\nu}{}^{\mu}{}_{\lambda} \mathcal{B}_{\mu}{}^{\nu}{}_{\alpha} \mathcal{B}_{\beta}{}^{\lambda}{}_{\gamma} \mathcal{A}_\delta
    + F_4 \mathcal{B}_{\alpha}{}^{\mu}{}_{\beta}\mathcal{B}_{\gamma}{}^{\nu}{}_{\delta}\mathcal{A}_\mu \mathcal{A}_\nu \bigg].
    \end{split}
\end{equation}
In the above action, the covariant derivative and the curvature tensor are defined
with respect to the symmetric part of the connection, meaning that $\nabla = \nabla^{\Gamma}$ 
and $\mathcal{R} = \mathcal{R}^{\Gamma}$. 

One of the most important features of this model is that the lack of a metric tensor  implies that the number of terms in the action is finite. This property is known as the \textit{rigidity} of the model.
The fact that all the coupling constants are dimensionless  suggests that the model is power-counting renormalizable, so that in the hypothetical scenario of its quantization all possible counter-terms should have the form of the ones already written in Eq. \eqref{PAG_action}.  This is something desirable from a stand point of Quantum Field Theory view. The reason is that the superficial degree of divergence vanishes. Additionally, the dimensionless nature of the coupling constants also suggests a conformal symmetry, at least at a classical level. An explicit implementation of this idea is likely to require an understanding of the projective invariance of the theory, along the lines of \cite{Olmo:2022ops}, though this aspect will not be explored in this paper. 

In the torsion-free sector, the field equations coming from the variation of the action turn out to be a generalization of the Einstein's vacuum field equations. In that sense, the space of solutions of GR in vacuum is a subspace of solutions of the Polynomial Affine Gravity. 
Finally, it is possible to couple a scalar field using the \textit{dimensional analysis} technique introduced above without the necessity of having a metric structure on the manifold. The kinetic terms would couple to a combination of tensors and tensor densities, while the potential term should rescale the volume element. ln the torsionless limit, the resulting equations turn out to recover Einsteins's theory coupled to a scalar field, see Ref.~\cite{castillofelisola2023inflationary}, giving support to this approach.

Since we are interested in the study of cosmology, we need to impose first the symmetries of the cosmological principle on 
the irreducible fields associated to the affine connection, which are $\Gamma$, $\mathcal{B}$ and $\mathcal{A}$. 
In order to build up the ansatz, we compute the Lie derivative of each irreducible field along the Killing vectors $\xi_i$ that generate the symmetries of homogeneity (translations) $\mathcal{P}_i$ and isotropy (rotations) $\mathcal{J}_i$. 
A derivation of the Killing vectors along with an explicit computation of the Lie derivative can be found 
in Ref. \cite{Castillo-Felisola17}. In what follows we shall briefly summarize the results.

The computation of the Lie derivative of $\Gamma$ along the Killing vectors determines its coefficients: 
\begin{eqnarray}
    \label{G_ansatz}
    \ctG{t}{t}{t} & =& f(t), \ \ \ \ \ \ctG{i}{t}{j}  = g(t) S_{i j}, \\
    \ctG{t}{i}{j} &=& h(t) \delta^{i}_{j} , \ \ \ \ctG{i}{j}{k}  = \ctg{i}{j}{k}, \nonumber
\end{eqnarray}
where $S_{ij}$ is a three-dimensional rank two symmetric tensor defined as 
\begin{equation}\label{eq:Sij}
    S_{i j}=\left(\begin{array}{ccc}
    \frac{1}{1-\kappa r^2} & 0 & 0 \\
    0 & r^2 & 0 \\
    0 & 0 & r^2 \sin ^2 \theta
    \end{array}\right),
\end{equation}
and $\gamma$ is the three-dimensional symmetric connection compatible with the desired symmetries, which can be written as
\begin{align*}
    \ctg{r}{r}{r} & = \frac{\kappa r}{1 - \kappa r^2}, & \ctg{\theta}{r}{\theta} & = \kappa r^3 - r, \\
    \ctg{\varphi}{r}{\varphi} & = \left(\kappa r^3 - r\right)\sin^2\theta, & \ctg{r}{\theta}{\theta} & = \frac{1}{r}, \\
    \ctg{\varphi}{\theta}{\varphi} & = -\cos\theta\sin\theta, & \ctg{r}{\varphi}{\varphi} & = \frac{1}{r}, \\
    \ctg{\theta}{\varphi}{\varphi} & = \frac{\cos \theta}{\sin \theta}.
\end{align*}
Additionally, the affine function $f(t)$ can be set to zero by a re-parametrization of the $t$ coordinate \cite{Castillo_Felisola_2022_EPJC}. 
Therefore, there are only two nontrivial functions to define completely the symmetric part of the connection.

Following a similar procedure for the torsion tensor, it is possible to determine the ansatz compatible with 
the required symmetries. For its traceless part $\mathcal{B}$, the non-vanishing components are
\begin{equation}
\label{B_ansatz}
\begin{aligned}
    \B{\theta}{r}{\varphi} & = \psi (t) r^2\sin\theta \sqrt{1 - \kappa r^2}, &
    \B{r}{\theta}{\varphi} & =\frac{\psi (t) \sin \theta}{\sqrt{1 - \kappa r^2}}, \\
    \B{r}{\varphi}{\theta} & =\frac{\psi(t)}{ \sqrt{1-\kappa r^{2}} \sin \theta},
\end{aligned}
\end{equation}
while the nontrivial component of its vectorial part $\mathcal{A}$ is
\begin{equation}
    \label{A_ansatz}
    \A{t} = \eta(t).
\end{equation}

Finally, the field equations are obtained by varying the action with respect to each irreducible field using Kijowski's 
formalism, see Ref. \cite{KJ_Formalism,Castillo_Felisola_2020}. The complete  set of field equations for each irreducible field can be found in Refs.~\cite{Castillo_Felisola_2020,Castillo-Felisola_2023}.

\section{Cosmological solutions}
\label{sec:solutions}
Varying the action (\ref{PAG_action}) and using the ansatz in Eqs. \eqref{G_ansatz}, \eqref{B_ansatz} and \eqref{A_ansatz}, the field equations become
\begin{equation}
    \label{Feq_1}
    \left(B_3\left(\dot{g} + gh + 2\kappa\right) - 2B_4\left(\dot{g} - gh\right) + 2D_6\eta g - 2F_3\psi^2\right)\psi = 0,
\end{equation}

\begin{equation}
    \label{Feq_2}
    (B_3\eta\psi -2B_4\eta\psi + C_1(\dot{\psi} - 2h\psi))g = 0,
\end{equation}

\begin{eqnarray}
    \label{Feq_3}
    \left(B_3 + 2B_4\right)\eta g\psi &+& 2C_1(\kappa\psi + 4gh\psi - g\dot{\psi} - \psi\dot{g}) \nonumber \\ &-& 2\psi^3(D_1 - 2D_2 + D_3) = 0,
\end{eqnarray}

\begin{eqnarray}
    \label{Feq_4}
    B_3(\eta(h\psi -\dot{\psi}) &-&\psi\dot{\eta}) - 2B_4(\eta(-h\psi - \dot{\psi}) - \psi\dot{\eta})   \\ &+& C_1(4h^2\psi + 2\psi\dot{h} -\ddot{\psi}) + D_6\eta^2\psi = 0 \ ,\nonumber 
\end{eqnarray}

\begin{eqnarray}
    \label{Feq_5}
    B_3(\dot{g} &+& gh + 2\kappa)\eta - 2B_4\left(\dot{g} - gh\right)\eta \nonumber \\ &+& C_1(2\kappa h + 4gh^2  + 2g\dot{h}- \ddot{g}) \nonumber \\ &-& 6h\psi^2\left(D_1 - 2D_2 + D_3\right) \nonumber \\ &+& D_6 \eta^2 g - 6F_3\eta\psi^2 = 0
\end{eqnarray}
where $\mathcal{A}$ and $\mathcal{B}$ yield only one equation each, namely,  Eq. \eqref{Feq_1} and Eq. \eqref{Feq_5}, respectively, 
while the remaining three come from $\Gamma$. Notice that we have four unknown functions 
of time ($g(t)$, $h(t)$, $\psi(t)$ and $\eta(t)$) while there are five differential equations, meaning that the system is overdetermined. We will show that, even so, the system can actually be solved analytically without any assumption. For this purpose, we have developed a \textit{logical scheme} that allows us to seek systematically the solutions by branches. First, notice that Eqs. \eqref{Feq_1} 
and \eqref{Feq_2} follow the form
\begin{dmath}
    \label{Feq_1_1}
    \mathcal{F}(g, \dot{g}, h,\psi,\eta)\psi = 0,
\end{dmath}
\begin{dmath}
    \label{Feq_2_2}
    \mathcal{G}(h,\psi, \dot{\psi}, \eta)g = 0,
\end{dmath}
where
\begin{dmath}
\mathcal{F}(g,\dot{g},h,\psi,\eta)  \equiv B_3\left(\dot{g} + gh + 2\kappa\right) - 2B_4\left(\dot{g} - gh\right) + 2D_6\eta g - 2F_3\psi^2.
\end{dmath}
\begin{dmath}
\mathcal{G}(h,\psi, \dot{\psi}, \eta)  \equiv B_3\eta\psi -2B_4\eta\psi + C_1\left(\dot{\psi} - 2h\psi\right).
\end{dmath}
Thus, using Eqs. \eqref{Feq_1_1} and \eqref{Feq_2_2} it is possible to distinguish four different branches:
\begin{itemize}
    \item {\bf First branch:}  { $ \ \ \ \mathcal{F}(g,h,\psi,\eta) \! = 0 \ \wedge \ \mathcal{G}(h,\psi,\eta) \!= \! 0$}.
    \item {\bf Second branch:} $ \mathcal{F}(g,h,\psi,\eta)  = 0 \ \wedge \ g  = 0$.
    \item {\bf Third branch:}   $\ \mathcal{G}(h,\psi,\eta)   = 0   \wedge  \psi  = 0$.
    \item {\bf Fourth branch:} $\psi  = 0  \wedge  g  = 0$.
\end{itemize}
The first branch is the most interesting one in terms of the richness of solutions, while branch four is the simplest one. 

\subsection{First branch}
\label{sec:first_branch}

This branch contains the most general case subject to $\mathcal{F}(g,h,\psi,\eta) = 0$ and $\mathcal{G}(h,\psi,\eta) = 0$, with the field equations being
\begin{dmath}
    \label{Feq_B1_1}
    B_3\left(\dot{g} + gh + 2\kappa\right) - 2B_4\left(\dot{g} - gh\right) + 2D_6\eta g - 2F_3\psi^2 = 0,
\end{dmath}
\begin{dmath}
    \label{Feq_B1_2}
    B_3\eta\psi -2B_4\eta\psi + C_1\left(\dot{\psi} - 2h\psi\right) = 0,
\end{dmath}
\begin{dmath}
    \label{Feq_B1_3}
    \left(B_3 + 2B_4\right)\eta g\psi + 2C_1\left(\kappa\psi + 4gh\psi - g\dot{\psi} - \psi\dot{g}\right) + 2\psi^3\left(2D_2 - D_1 - D_3\right) = 0,
\end{dmath}
\begin{dmath}
    \label{Feq_B1_4}
    B_3\left(\eta\left(h\psi - \dot{\psi}\right) -\psi\dot{\eta}\right) - 2B_4\left(\eta\left(-h\psi - \dot{\psi}\right) -\psi\dot{\eta}\right) 
    + C_1\left(4h^2\psi + 2\psi\dot{h} -\ddot{\psi}\right) + D_6\eta^2\psi = 0,
\end{dmath}
\begin{dmath}
    \label{Feq_B1_5}
    B_3\left(\dot{g} + gh + 2\kappa\right)\eta - 2B_4\left(\dot{g} - gh\right)\eta + C_1\left(2\kappa h + 4gh^2 + 2g\dot{h} - \ddot{g}\right) +
    6h\psi^2\left(2D_2 - D_1 - D_3\right) + D_6 \eta^2 g - 6F_3\eta\psi^2 = 0.
\end{dmath}
To find an expression for $\eta(t)$, one must solve Eq.~\eqref{Feq_B1_2} 
\begin{equation}
    \label{B1_eta}
    \eta(t) = \left(\frac{2h\psi - \dot{\psi}}{\psi}\right)\left(\frac{C_1}{B_3 - 2B_4}\right).
\end{equation}
Replacing the above expression for $\eta(t)$ into Eq. \eqref{Feq_B1_4}, leads to two sub-branches for the $h(t)$ function, namely
\begin{align}
    \label{B1_h}
    h_I(t) & = \frac{\dot{\psi}}{2\psi} & & \wedge &  h_{II}(t) & = \frac{\dot{\psi}}{\psi}\left(\frac{C_1 D_6}{3B_3^2 - 8B_3B_4 + B_4^2 + 2C_1D_6}\right).
\end{align}
Using $h_I(t)$, then Eq. \eqref{Feq_B1_3} turns into
\begin{equation}
    -\left(D_1 - 2D_2 + D_3\right)\psi^3 + C_1\left(\psi\left(\kappa - \dot{g}\right) + g\dot{\psi}\right) = 0,
\end{equation}
which allows us to solve $g(t)$ in terms of the $\psi(t)$ 
\begin{equation}
    \label{B1_g}
    g(t) = \psi(t) \left(g_0 + \int_1^t \left(\frac{\kappa}{\psi(\tau)} - \psi(\tau) \left(\frac{D_1 - 2D_2 + D_3}{C_1}\right)\right) \mathrm{d}\tau\right),
\end{equation}
where $g_0$ is an integration constant. The above expression also solves Eq. \eqref{Feq_B1_5}. Then, Eq. \eqref{Feq_B1_1} becomes a first-order integro-differential equation 
\begin{dmath}
    \label{psi_integro_diff_equation}
    \dot{\psi}\left(g_0 + \int_1^t \left(\frac{\kappa}{\psi(\tau)} - \psi(\tau) \alpha\right) \mathrm{d}\tau\right)\beta -
    \psi^2 \gamma + 2\kappa\beta = 0.
\end{dmath}
where $\alpha$, $\beta$ and $\gamma$ are related to the coupling constants
by the following relation
\begin{align}
    \alpha & = \left(\frac{D_1 - 2D_2 + D_3}{C_1}\right), \\
    \beta & = \left(\frac{3B_3 - 2B_4}{2}\right), \\
    \gamma & = \left(\beta - 2B_3\right)\alpha + 2F_3,
\end{align}
For the special case $\kappa = 0$ and defining $\psi (t) \equiv \dot{\phi}(t)$, Eq.~\eqref{psi_integro_diff_equation} can be turned into a second-order differential equation for $\phi(t)$ of the form
\begin{dmath}
\label{diff_eq_phi}
    \ddot{\phi}\left(g_0  - \phi\alpha\right)\beta - \dot{\phi}^2 \gamma  = 0,
\end{dmath}
whose solution is
\begin{equation}
    \label{phi_general}
    \phi(t) =\frac{g_0}{\alpha} + \lambda\left(t -t_0\right)^{\frac{\alpha\beta}{\alpha\beta + \gamma}},
\end{equation}
where $\lambda $ and $t_0$ are integration constants. Using the above solution, it is straightforward to recover the original function
\begin{equation}
    \psi(t) =\frac{ \lambda\alpha\beta}{\alpha\beta + \gamma}\left(t -t_0\right)^{-\frac{\gamma}{\alpha\beta + \gamma}}.
\end{equation}
Using the above expression and from the relations in Eqs. \eqref{B1_eta}, \eqref{B1_h} and \eqref{B1_g} it is direct to find the rest of the
affine functions
\begin{dmath}
\eta(t)  = 0
\end{dmath}
\begin{dmath}
h(t) =  -\frac{\gamma}{2\left(\alpha\beta + \gamma\right)\left(t - t_0\right)}
\end{dmath}
\begin{dmath}
g(t) =g_1\left(t - t_0\right)^{-\frac{\gamma}{\alpha\beta + \gamma}}
-\frac{\alpha^2\beta\lambda^2 }{\alpha\beta + \gamma}\left(t - t_0\right)^{\frac{\alpha\beta-\gamma}{\alpha\beta + \gamma}}\end{dmath}
where $g_1$ is an integration constant.

Finally, from Eq. \eqref{B1_h} the second branch of $h(t)$ leads to a similar first order integro-differential equation 
which can not be solved analytically, which is why, for the moment we will exclude in this paper.

\subsection{Second branch}
\label{sec:second_branch}

The second branch imposes $\mathcal{F}(g,h,\psi,\eta)  = 0$ and $g(t) = 0$ restrictions which leads to the following field equations:
\begin{dmath}
    \label{Feq_B2_1}
    \kappa B_3 \psi - F_3\psi^3= 0,
\end{dmath}
\begin{dmath}
    \label{Feq_B2_3}
    \kappa C_1\psi - \psi^3 D = 0 ,
\end{dmath}
\begin{eqnarray}
    \label{Feq_B2_4}
    B_3(\eta(h\psi -\dot{\psi}) &-&\psi\dot{\eta}) - 2B_4(\eta(-h\psi - \dot{\psi}) - \psi\dot{\eta})   \\ &+& C_1(4h^2\psi + 2\psi\dot{h} -\ddot{\psi}) + D_6\eta^2\psi = 0 \ ,\nonumber 
\end{eqnarray}
\begin{eqnarray}
    \label{Feq_B2_5}
    B_3\kappa\eta + C_1\kappa h - 3h\psi^2 D - 3F_3\eta\psi^2 = 0,
\end{eqnarray}
where $D = D_1 - 2D_2 + D_3$.

From Eq. \eqref{Feq_B2_1} it is possible to find an expression for $\psi(t)$ in the form
\begin{equation}
	\label{B2_Psi}
    \psi(t) = \pm \sqrt{\frac{\kappa B_3}{F_3}} .
\end{equation}
Using the compatibility condition from Eq. \eqref{Feq_B2_3}, leads to a relation between
the coupling constant
\begin{equation}
 C_1 F_3 = D B_3.
\end{equation}
Taking the algebraic expression for $C_1$ and replacing Eq. \eqref{B2_Psi} in Eq.~\eqref{Feq_B2_5} leads to a relation between $h(t)$ and $\eta(t)$ of the form
\begin{equation}
    h(t) = - \eta(t)\frac{F_3}{D}
\end{equation}
Combining the above result along with Eq. \eqref{B2_Psi} turns Eq. \eqref{Feq_B2_4} into a first order differential equation of the form
\begin{equation}
 \dot{\eta} - \eta^2\left(\frac{D_6}{3B_3 - 2B_4} + \frac{F_3}{D}\right) = 0
\end{equation}
whose solution is
\begin{equation}
    \eta(t) = \frac{D\left(2B_4 - 3B_3\right)}{\left(D \eta_0 + tF_3\right)\left(3B_3 - 2B_4\right) + DD_6 t}.
\end{equation}
Then, it is straightforward to obtain $h(t)$ 
\begin{equation}
    h(t) = \frac{F_3\left(2B_4 - 3B_3\right)}{\left(D \eta_0 + tF_3\right)\left(3B_3 - 2B_4\right) + DD_6 t}
\end{equation}
It is worth to remark that the above solutions is only valid for the special case 
where $\kappa \neq 0$, this can be seen directly from Eqs. \eqref{Feq_B2_1} and \eqref{Feq_B2_3}.

If we impose the constraint $\kappa = 0$, then Eq. \eqref{Feq_B2_1} tells us that $\psi(t) = 0$ solves entirely the other
equations, and the remaining functions $h(t)$ and $\eta(t)$ are unknown.

\subsection{Third branch}

The constraint $\mathcal{G}(h,\psi,\eta) = 0$ and $\psi(t)=0$ imposes a strong restriction to the system of differential equations, which boils down to 
\begin{dmath}
    g\eta^2D_6 + 2B_4\eta \left(gh - \dot{g}\right) + B_3\eta\left(2\kappa + gh + \dot{g}\right) + 
    C_1\left(2h\left(\kappa + 2gh\right) + 2g\dot{h} - \ddot{g}\right) = 0.
\end{dmath}
The above differential equation has three unknown functions of time $h(t)$, $g(t)$ and $\eta(t)$ which cannot be solved without 
further restriction, or by providing an ansatz for two functions.

\subsection{Fourth branch}

In the fourth branch defined by $g(t) = 0$ and $\psi (t) = 0$, the field equations are reduced
to a single algebraic equation
\begin{equation}
    \kappa\left(hC_1 + B_3\eta\right) = 0,
\end{equation}
the above equation can be solved by setting $\kappa = 0$, then, the functions $h(t)$ and $\eta(t)$ are undetermined,
or by the relation $\eta(h) = - h(t)\left(\frac{C_1}{B_3}\right)$, where $h(t)$ is an arbitrary function.

\section{Analysis of the solutions}
\label{sec:analysis}

There are only two nontrivial and analytic solutions to the field equations, which were found
in Sec. \ref{sec:first_branch} and \ref{sec:second_branch}. Nonetheless, 
we will restrict ourselves to the analysis of the first branch \ref{sec:first_branch} for reasons that will become clear in the next subsection.

In order to simplify our calculations, without loss of generality we shall set up the integration constant $t_0 \to 0$ because of the time shift symmetry of the equations. We also introduce the constant factor $\Omega$ as
\begin{equation}\label{eq:Omega}
\Omega \equiv \frac{\gamma}{\alpha\beta},
\end{equation}
along with the re-scalings by $\lambda$
\begin{align}
\psi(t) \to \frac{\psi(t)}{\lambda}, && g(t) \to \frac{g(t)}{\lambda^2},
\end{align}
which allows us to write 
\begin{align}
	h(t) & = -\frac{\Omega}{2t\left(1 + \Omega\right)}, \label{sol_h}\\
	g(t) & = g_1 t^{-\frac{\Omega}{1 + \Omega}} - \left(\frac{\alpha}{1 + \Omega}\right) t^{\frac{1 - \Omega}{1 + \Omega}}, \label{sol_g} \\
	\psi(t) & =\left(\frac{1}{1 + \Omega}\right)t^{-\frac{\Omega}{1 + \Omega}},  \label{sol_p}\\ 
	\eta(t) & = 0, \label{sol_n}
\end{align}
where $g_1$ is actually the original $g_1$ divided by $\lambda^2$. In this representation, the parameter $\Omega$ determines the time dependence, while the ratio $g_1/\alpha$ characterizes the relative amplitude of the two terms in (\ref{sol_g}). Regarding the time dependence of (\ref{sol_g}), note that if $\Omega>1$, then the two terms diverge at the origin of time, being the first one dominant at the beginning but decaying faster than the second. If $0<\Omega<1$, only the term proportional to $g_1$ diverges at $t=0$. For $-1<\Omega<0$, this term vanishes at $t=0$ and the second one is dominant at late times. If $\Omega<-1$  then both terms diverge at $t=0$. It is easy to see that  $\psi(t)$ is regular at the origin of time only when $0<\Omega<1$. Thus, the only way to make $\psi(t)$ and $g(t)$ regular at $t=0$ is by setting the integration constant $g_1$ to zero when $0<\Omega<1$. Regardless of $\Omega$, the function $h(t)$ always diverges as $t\to 0$.

\subsection{Emergent metrics}
We now need to interpret the physical implications that follow from the above solutions to the connection field equations. The first step in this direction is to explore the descendant metric structures that emerge from the affine connection. In this sense, it is convenient to recall the definition of a metric tensor:

\textbf{Definition}. Let $\mathcal{M}$ be a smooth manifold of dimension $n$. At each
point $p \in \mathcal{M}$ there is a vector space $T_p\mathcal{M}$, called the tangent
space. A metric tensor at the point $p$ is a function $g_p (X_p, Y_p)$ which takes as 
inputs a pair of tangent vectors $X_p$ and $Y_p$ at $p$ and produces a real number, which
should vary smoothly as one changes the point, so that the following conditions are 
satisfied: (i) $g$ is bilinear, meaning that it is linear separately in each argument, 
(ii) $g$ is symmetric provided that for all vector $X_p$ and $Y_p$ we have $g_p(X_p, Y_p) = g_p (Y_p, X_p)$, 
(iii) $g$ is non-degenerate and, therefore, it can be inverted. 

Although all three points are relevant in order to define a metric tensor, the last point
plays a crucial part, because it ensures the existence of the inverse of $g_{\mu\nu}$ defined by $g^{\mu\nu}$, and thus,
it is possible to define the notion of distance. Additionally, for our purposes, the metric should define a causal structure, which is 
equivalent to a constraint in its signature. This is a highly nontrivial point that will 
constrain the parameters of the theory.

Following the literature, we can consider (at least) two notions of metric that yield nontrivial results in the scenario that we are considering. Up to a proportionality constant, the first emergent metric can be identified with the symmetrized Ricci tensor $\mathcal{R}_{(\mu\nu)}$, defined by the contraction of the Riemann tensor,\footnote{The Riemann tensor is defined by the commutator of covariant derivatives acting on a vector, therefore, it does not required the existence of a metric structure.}
\begin{equation}
    \label{chain}
    \Ri{\beta\delta} = \Rie{\alpha\beta}{\alpha}{\delta} .
\end{equation}
A second metric structure, comes from the contraction of the product of two torsion tensors.\footnote{This comes from the antisymmetric part of the affine connection} This idea was first introduced by Poplawski in Ref. \cite{Pop_awski_2013}, and the metric structure is defined as follows
\begin{equation}
    \mathcal{P}_{\alpha\delta} = \left(\B{\alpha}{\beta}{\gamma} + \delta^{\beta}_{[\gamma}\A{\alpha]}\right)\left(\B{\beta}{\gamma}{\delta} + \delta^{\gamma}_{[\delta}\A{\beta]}\right).
\end{equation}
Using the cosmological ansatz presented in Eq. \eqref{G_ansatz}, the nonzero components of the Ricci tensor become
\begin{align}
    \label{Ricci_tensor}
    \mathcal{R}_{tt} & = -3\left( \dot{h} + h^2\right), & \mathcal{R}_{rr} & = \frac{\dot{g} + gh + 2\kappa}{1 - \kappa r^2},
\end{align}
and $\mathcal{R}_{\theta\theta}$ and $\mathcal{R}_{\phi\phi}$ are related to $\mathcal{R}_{rr}$ as one would expect. Poplawski's metric computed using Eqs. \eqref{B_ansatz} and \eqref{A_ansatz} becomes
\begin{align}
    \label{Pop_tensor}
    \mathcal{P}_{tt} & = \eta^2, & \mathcal{P}_{rr} & = -\frac{2\psi^2}{1 - \kappa r^2},
\end{align}
and $\mathcal{P}_{\theta\theta}$ and $\mathcal{P}_{\phi\phi}$ are related to $\mathcal{P}_{rr}$. Note that the two prescriptions above for an emergent metric do not coincide, as one would expect, because the former is generated by the symmetric part of the affine connection, whereas the latter is defined by its antisymmetric part. We now proceed to interpret if those results make any physical sense. A minimal requirement is that the metric tensor be invertible. Given that in our case of interest $\eta(t)=0$, we can readily see that the Poplawski tensor is not a suitable metric. On the contrary, the expressions (\ref{Ricci_tensor}) evaluated on the solutions (\ref{sol_h}) and (\ref{sol_g}) yield nonzero results, which allows us to proceed further. If this Ricci tensor is to be identified with an homogenous and isotropic metric, we should have $g_{tt} =  \mathcal{R}_{tt}/\mathcal{R}_0 $ and $a^2(t)S_{ij}=  \mathcal{R}_{ij}/\mathcal{R}_0$, where $S_{ij}$ was defined in (\ref{eq:Sij}), and $\mathcal{R}_0$ is some constant with dimensions of curvature used to get a dimensionless metric.

Using Eqs. \eqref{sol_h}, \eqref{sol_g} and \eqref{Ricci_tensor} we compute the components 
of the Ricci tensor (for $\kappa=0$)
\begin{align}
    \mathcal{R}_{tt} & = -\frac{3\Omega\left(2 + 3 \Omega\right)}{4t^2 \left(1 + \Omega\right)^2}, \label{R_temporal} \\
    \mathcal{R}_{rr} & =\frac{\alpha\left(3\Omega - 2\right)}{2\left(\Omega + 1\right)^2}t^{-\frac{2\Omega}{1 + \Omega}} \label{R_spatial}.
\end{align}
Notice that if $\Omega \neq -1$ then the Ricci tensor can indeed act as a metric tensor because it defines an invertible matrix. Now we need to check if this object may indeed satisfy the Lorentzian signature constraint. Obviously, the theory should be signature invariant in the sense that one is free to choose between $-+++$ or $+---$. This may impose restrictions on the model parameters that must be taken into account. To proceed, we must note that for the choice $-+++$ we should have the constraints
\begin{eqnarray}
\Omega\left(2 + 3 \Omega\right)&>&0 \nonumber \\
\alpha\left(3\Omega - 2\right)&>&0 \nonumber  \ ,
\end{eqnarray}
which are equivalent to demanding that $\mathcal{R}_{tt}<0$ and $\mathcal{R}_{rr} >0$. On the other hand, for $+---$ the corresponding conditions turn into
\begin{eqnarray}
\Omega\left(2 + 3 \Omega\right)&<&0 \nonumber \\
\alpha\left(3\Omega - 2\right)&<&0 \nonumber  \ .
\end{eqnarray}
One can check that the area of the rectangle in the plane $\alpha, \Omega$ that allows for the free choice of signature has vanishing area. Since it seems desirable to have a set of solutions which generate an emergent metric such that its associated physics is invariant under signature changes, it is worth scrutinizing this point more carefully. In this sense, the proportionality constant $\mathcal{R}_0$ that we introduced above to relate the effective metric and the symmetric part of the Ricci tensor can be chosen such that these two sectors of the theory do overlap. In fact, since $\Omega$ is an effective parameter that depends on the coefficients that specify the Lagrangian of the theory, it is natural to admit that $\mathcal{R}_0$ could be dependent on such parameters as well. Thus, if we set $\mathcal{R}^{-}_0>0$ for the branch with signature $-+++$, but take $\mathcal{R}^+_0=-\mathcal{R}^{-}_0$ in the branch $+---$, then the two branches would lead exactly to the same signature. Obviously, had we taken $\mathcal{R}^{+}_0>0$ for the  $+---$ branch and set  $\mathcal{R}^-_0=-\mathcal{R}^{+}_0$, the resulting metric would have the opposite signature for both branches. We thus conclude that the arbitrary proportionality constant $\mathcal{R}_0$ can be used to impose the freedom we required in the choice of signature. We will now check explicitly that this reasoning is consistent by considering explicitly the solutions corresponding to the two possible signatures.
Let us first study the $-+++$  case, for which we set $\mathcal{R}^{-}_0=1$ for simplicity. Before diving into the analysis, it is convenient to write the above expression using the cosmic time coordinate $\tau$. This can be achieved by the following rule
\begin{equation}
-\frac{3\Omega\left(2 + 3 \Omega\right)}{4t^2 \left(1 + \Omega\right)^2} \mathrm{d}t^2 = - \mathrm{d}\tau^2 \to t(\tau) = t_0e^{\pm\frac{2|1 + \Omega|\tau}{\sqrt{3\Omega(2 + 3\Omega)}} },
\end{equation}
where $t_0$ is an irrelevant integration constant. From the above relation we rewrite the spatial part of the Ricci tensor and by taking its square root leads to the affine scale factor
\begin{equation}
a(\tau) = e^{\mp\frac{2\tau\Omega}{\sqrt{3\Omega(2 + 3\Omega)}}}\sqrt{\frac{\alpha(3\Omega - 2)}{2(1 + \Omega)^2}}.
\end{equation}
Using the above expression it is straightforward to derive the Hubble function as
\begin{equation}
H(\tau)\equiv \frac{a_\tau}{a} = \mp\frac{2\Omega}{\sqrt{3\Omega(2 + 3\Omega)}}.
\end{equation}
Notice that the scale factor depends on the values of the parameters $\alpha$ and $\Omega$, whereas the {Hubble factor} depends only on the value
of $\Omega$. This Hubble function describes a contracting/expanding de Sitter universe if $\Omega>0$.  For the $-+++$  signature we are considering, the case $\alpha>0$ requires $\Omega>2/3$ and the resulting universe is indeed contracting/expanding. For $\alpha<0$, we have two possible intervals for $\Omega$, namely, 
$\Omega<-2/3$ and $0<\Omega<2/3$. In the latter case, the solutions are also contracting/expanding, but in the former, they represent an expanding/contracting de Sitter universe. 

If we consider now the $+---$ signature, with $\mathcal{R}^{+}_0=1$, we find that the time coordinates are related by the same expressions as before but with $\Omega(2 + 3\Omega)$ replaced by its modulus. The same happens with the scale factor and the Hubble function, which also require taking the modulus of $\alpha(3\Omega - 2)$. Thus, the functional expressions are basically the same, though defined in different intervals of the parameter domain. These domains can be identified with the choice of proportionality constant $\mathcal{R}^{\pm}_0$ that we proposed above.

\subsection{Special cases}

Although Eq. \eqref{phi_general} is the general solution to Eq. \eqref{diff_eq_phi}, there are some special cases that arise when  $\alpha = 0$ and $\alpha\beta + \gamma = 0$ that must be studied separately. We will now address those cases. 

When $\alpha = 0$, the differential equation \eqref{diff_eq_phi} simplifies to 
\begin{dmath}
\label{diff_eq_phi_alpha_0}
    \ddot{\phi}g_0\beta - \dot{\phi}^2 \gamma  = 0,
\end{dmath}
and its solution can be written as
\begin{equation}
\phi(t) = \phi_0 + \frac{\beta g_0}{\gamma}\log\left(\gamma(t-t_0)\right),
\end{equation}
where $\phi_0$ and $t_0$ are integration constants. From this, it is trivial to find the other functions
\begin{align}
\label{special_first_solution}
\psi(t) & = \frac{\beta g_0}{2\gamma(t-t_0)} & h(t) & = -\frac{\gamma}{2\gamma(t-t_0)} \\
g(t) & = \frac{g_1}{\gamma(t-t_0)}
\end{align}
where $g_1$ is another integration constant. Just like before, we can set  $t_0 \to 0$ without loss of generality. Computing the Ricci tensor using Eqs. \eqref{Ricci_tensor}, we are led to 
\begin{align}
\label{Ricci_special_1}
\mathcal{R}_{tt} & = -\frac{9}{4t^2} & \mathcal{R}_{rr} & = -\frac{3g_1}{2t^2\gamma}.
\end{align}
Notice that, the signature of the Ricci tensor is already fixed to be $+ - - -$, though it can be inverted by a suitable choice of $\mathcal{R}_0$. From the above expression we can find the relation between the time coordinate $t$ and the cosmic time $\tau$ as follows
\begin{equation}
\label{special_1_t}
t(\tau) = t_1e^{\pm\frac{2\tau}{\sqrt{3}}},
\end{equation}
where $t_1$ is an integration constant, which determines the amplitude of the function at $\tau=0$. By using the above result in Eq. \eqref{Ricci_special_1},  the scale factor can be identified as 
\begin{equation}
a(\tau) =  \sqrt{\frac{3g_1}{2\gamma}}e^{\mp\frac{2\tau}{\sqrt{3}}} \ ,
\end{equation}
which leads to the Hubble function
\begin{equation}
H(\tau) = \mp\frac{2}{\sqrt{3}}.
\end{equation}
The above scale factor describes a contracting/expanding de Sitter universe.

The second case is $\alpha\beta + \gamma = 0$ and leads to
\begin{dmath}
\label{diff_eq_phi_alpha_beta_gamma_0}
    \ddot{\phi}\left(g_0 - \alpha\phi\right) + \dot{\phi}^2 \alpha  = 0,
\end{dmath}
whose solution is
\begin{equation}   
\phi(t) = \phi_0 e^{\alpha\phi_1\left(t -t_0\right)} + \frac{g_0}{\alpha},
\end{equation}
where $\phi_0$ and $t_0$ are integration constants. Simple algebra allows us to recover the affine functions as 
\begin{align}
\label{special_second_solution}
\psi(t) & = \psi_0 e^{\frac{\left(t -t_0\right)}{\tau_0}} & h(t) & = \frac{\alpha }{2} \\
g(t) & =\psi(t)\left(g_1 - \frac{\psi(t)}{\phi_1}\right)  \nonumber
\end{align}
where $\psi_0 = \alpha\phi_0$ and $\tau_0^{-1} = \alpha\phi_1$ and $g_1$ is an integration constant. Setting $t_0$ to zero, the Ricci tensor becomes
\begin{align}
\mathcal{R}_{tt} & = -\frac{3}{4}\alpha^2 & \mathcal{R}_{rr} & = \frac{\alpha\psi}{2} \left[g_1(1+2\phi_1)-\frac{\psi}{\phi_1}\left(1+4\phi_1\right)\right] 
\end{align}
which has as natural signature $-+++$. Working on the cosmic time implies a simple transformation
\begin{equation}
t(\tau) =\pm \frac{\tau}{\tau_0}\sqrt{\frac{4}{3}},
\end{equation}
which is a linear rescaling of the time coordinate. Then, the scale factor squared is written as
\begin{equation}
a^2(\tau) = Ae^{\mp\frac{\tau}{\tau_1}} - Be^{\mp\frac{2\tau}{\tau_1}},
\end{equation}
where we have defined the following constant
\begin{align}
\label{definitions}
\tau_1 & = \tau_0^2 \sqrt{\frac{3}{4}} & A & = \psi_0g_1\left(\frac{\alpha\tau_0 + 1}{\tau_0}\right) & B & = \psi_0^2\left(\frac{\alpha\tau_0 + 2}{\tau_0\phi_1}\right)
\end{align}
and the Hubble function take form of
\begin{equation}
H(\tau) = \mp \frac{1}{2\tau_1} \mp \frac{B}{2\tau_0\left(B - Ae^{\pm \frac{\tau}{\tau_1}}\right)}
\end{equation}
It is evident that this Hubble function (not constant) and expansion factor (not a pure exponential) do not describe a de Sitter universe. Nonetheless, depending on the model parameters, one can find different interesting configurations, including asymptotically de Sitter universes. %This happens, for instance, by taking the positive sign in the expression for the Hubble function and taking the limit $\tau \to \infty$ the Hubble functions is a positive constant aslong as $\tau_1 > 0$ (which is always true by its definition in Eq. \eqref{definitions}) and $\tau_0 > 0$, and therefore, it is possible to have asymptotically de Sitter. On the other hand, by taking the negative sign on the Hubble function and the limit $\tau \to \infty$, the second term vanishes whereas the first term can provide for a negative Hubble.

\section{Final remarks}\label{sec:final_remarks}
In this work, we have analyzed homogeneous and isotropic (cosmological) solutions of a 
Polynomial Affine Gravity model in which torsion effects are explicitly considered, generalizing in this way previous results of Refs.~\cite{castillofelisola2019cosmological,Castillo_Felisola_2020}. In those works, it was shown that in the torsionless limit there are model parameters that recover vacuum cosmological solutions coupled with a cosmological constant, which allows to inteprete the Ricci tensor as an effective (connection-descendent) metric tensor, see Ref. \cite{Castillo_Felisola_2020,castillofelisola2023inflationary}.

Our introduction of torsion via the tensors \(\mathcal{A}\) and \(\mathcal{B}\) leads to the system of differential equations~\eqref{Feq_1}-\eqref{Feq_5}, from which we have identified four different branches of solutions. Although the third and fourth branches can not be solved exactly, because the system is under-determined, the first and second branches do admit an analytic treatment.

The second branch, which is more restrictive than the general case, has four differential equations that can be solved exactly, providing an expression for the functions $h(t)$ and $\eta(t)$. However, it is not possible to endow such solutions with a physical meaning because our two candidate prescriptions for an effective metric tensor turn out to be degenerate. A notion of distance can not be defined in that case. 

The first branch, which is the most general scenario, has five differential equations that can be
manipulated to yield a fundamental integro-differential equation, see Eq. \eqref{A_ansatz}, or a second-order differential equation, see Eq. \eqref{diff_eq_phi}. Exact solutions can be found for generic values of the parameters, see Eqs. \eqref{sol_h} and \eqref{sol_g}, and we also identified two singular cases that require a separate analysis. These are the second solution in Eq. \eqref{special_first_solution} and the third solution in Eq. \eqref{special_second_solution}.

Once the solutions were found, we checked the consistency of two potential definitions for an emergent metric tensor. The first metric is basically identified with the Ricci tensor (up to rescaling by a suitable parameter $\mathcal{R}_0$ to make it dimensionless) which is defined only by the symmetric part of the connection $\Gamma$. The second comes from the product of two torsion tensor. Nonetheless, we saw that the latter is not a suitable effective metric because the vectorial part of the torsion vanishes, $\eta (t) = 0$, leading in that way to a degenerate tensor.

The aforementioned metrics are defined (when possible) on the space of solutions. Their characteristic properties, like signature for example, are determined by the values of the couplings of the model, allowing changes of their values in different patches of the parameter space of the polynomial affine model of gravity. On physical grounds, one expects the signature to be Lorentzian, and we showed that its character depends on an arbitrary constant $\mathcal{R}_0$. 

The first solution, Eqs. \eqref{sol_h} and \eqref{sol_g} defines a well behaved Ricci tensor and, according to our criteria, it can be used as a metric tensor. Besides being non-degenerate, we noted that it is essential to reproduce a Lorentzian signature. Moreover, we argued that there is enough freedom in the definition of  this effective metric to make the choices $ - + + +$ and $+ - - -$ physically equivalent. We found that the scale factor describes a contracting/expanding de~Sitter universe for both signatures. A similar behaviour is found for the second solution in Eqs. \eqref{special_first_solution}.

The third solution, presented in Eq. \eqref{special_second_solution}, naturally selects the signature $ - + + +$ but can be flipped by a suitable choice of $\mathcal{R}_0$. In this case, the scale factor squared is a linear combination of two real exponential functions, departing from the de~Sitter behavior found in previous cases. From the perspective of GR, such a solution would not describe a vacuum universe. 

To conclude, we would like to highlight that  the introduction of torsion in Polynomial Affine Gravity has allowed us to identify an effective non-degenerate metric tensor with Lorentzian signature in the space of solutions. This facilitated the interpretation of solutions, as one could identify a cosmological time and use it to interpret the scale factor, which we found to represent de~Sitter and non-de~Sitter cosmologies.

\section{Acknowledgments}

We are specially thankful to the developers and maintainers of SageMath \cite{sagemath}, SageManifolds \cite{Gourgoulhon_2015,Gourgoulhon_2018}, and Cadabra \cite{peeters2018introducing,Peeters2018,Peeters_2007}. Those softwares were used extensively in our calculations. This work is supported by the Spanish Agencia Estatal de Investigaci\'on grant PID2020-116567GB-C21, funded by MCIN/AEI/10.13039/501100011033, FEDER, UE, and ERDF A way of making Europe, by the project PROMETEO/2020/079 (Generalitat Valenciana), and is based upon work from COST Action CA21136, supported by COST (European Cooperation in Science and Technology). OCF acknowledge the financial support received by ANID PIA/APOYO~AFB230003 (Chile) and FONDECYT Regular No.~1230110 (Chile).

% \bibliographystyle{spphys}
% \bibliography{References}

\begin{thebibliography}{10}
\providecommand{\url}[1]{{#1}}
\providecommand{\urlprefix}{URL }
\expandafter\ifx\csname urlstyle\endcsname\relax
  \providecommand{\doi}[1]{DOI \discretionary{}{}{}#1}\else
  \providecommand{\doi}{DOI \discretionary{}{}{}\begingroup
  \urlstyle{rm}\Url}\fi

\bibitem{Will_2014}
C.M. Will, Living Reviews in Relativity \textbf{17}(1) (2014).
\newblock \doi{10.12942/lrr-2014-4}

\bibitem{will_2018}
C.M. Will, \emph{Theory and Experiment in Gravitational Physics}, 2nd edn.
  (Cambridge University Press, 2018).
\newblock \doi{10.1017/9781316338612}

\bibitem{Weinberg:2008zzc}
S.~Weinberg, \emph{{Cosmology}} (2008)

\bibitem{Abbott_2016}
B.P. Abbott, et~al., Physical Review Letters \textbf{116}(24) (2016).
\newblock \doi{10.1103/physrevlett.116.241103}

\bibitem{Abbott_2017}
B.P. Abbott, et~al., The Astrophysical Journal \textbf{848}(2), L13 (2017).
\newblock \doi{10.3847/2041-8213/aa920c}

\bibitem{Einstein_GR}
A.~{Einstein}, Annalen der Physik \textbf{354}(7), 769 (1916).
\newblock \doi{10.1002/andp.19163540702}

\bibitem{Einstein_GR_Bases}
A.~{Einstein}, Sitzungsberichte der K\"oniglich Preussischen Akademie der
  Wissenschaften pp. 778--786 (1915)

\bibitem{Frie}
A.~Friedman, Zeitschrift f{\"u}r Physik \textbf{10}(1), 377 (1922).
\newblock \doi{10.1007/BF01332580}

\bibitem{Friedmann}
A.~{Friedmann}, Zeitschrift fur Physik \textbf{21}(1), 326 (1924).
\newblock \doi{10.1007/BF01328280}

\bibitem{Lema}
G.~{Lemaitre}, Monthly Notices of the Royal Astronomical Society \textbf{91},
  483 (1931).
\newblock \doi{10.1093/mnras/91.5.483}

\bibitem{Lema_Expansion}
G.~Lemaitre, Annales de la Société Scientifique de Bruxelles \textbf{53}, 51
  (1933)

\bibitem{Robertson_1}
H.P. Robertson, The Astrophysical Journal \textbf{82}, 284 (1935).
\newblock \doi{10.1086/143681}

\bibitem{Robertson_2}
H.P. Robertson, The Astrophysical Journal \textbf{83}, 187 (1936).
\newblock \doi{10.1086/143716}

\bibitem{Robertson_3}
H.P. Robertson, The Astrophysical Journal \textbf{83}, 257 (1936).
\newblock \doi{10.1086/143726}

\bibitem{Einstein_GR_Feqs}
A.~{Einstein}, Sitzungsberichte der K\"oniglich Preussischen Akademie der
  Wissenschaften pp. 844--847 (1915)

\bibitem{Einstein_Cosmological_Constant}
A.~{Einstein}, Sitzungsberichte der K\"oniglich Preussischen Akademie der
  Wissenschaften pp. 142--152 (1917)

\bibitem{PhysRev.160.1113}
B.S. DeWitt, Phys. Rev. \textbf{160}, 1113 (1967).
\newblock \doi{10.1103/PhysRev.160.1113}

\bibitem{PhysRev.162.1195}
B.S. DeWitt, Phys. Rev. \textbf{162}, 1195 (1967).
\newblock \doi{10.1103/PhysRev.162.1195}

\bibitem{PhysRevD.10.401}
S.~Deser, P.~van Nieuwenhuizen, Phys. Rev. D \textbf{10}, 401 (1974).
\newblock \doi{10.1103/PhysRevD.10.401}

\bibitem{PhysRevD.10.411}
S.~Deser, P.~van Nieuwenhuizen, Phys. Rev. D \textbf{10}, 411 (1974).
\newblock \doi{10.1103/PhysRevD.10.411}

\bibitem{Rotation_Curve}
V.C. {Rubin}, W.K. {Ford Jr.}, The Astrophysical Journal \textbf{159}, 379
  (1970).
\newblock \doi{10.1086/150317}

\bibitem{Rotation_Curve_2}
Y.~Sofue, V.~Rubin, Annual Review of Astronomy and Astrophysics \textbf{39}(1),
  137 (2001).
\newblock \doi{10.1146/annurev.astro.39.1.137}

\bibitem{NASERI2021100888}
M.~Naseri, J.T. Firouzjaee, Physics of the Dark Universe \textbf{34}, 100888
  (2021).
\newblock \doi{https://doi.org/10.1016/j.dark.2021.100888}

\bibitem{10.1093/mnras/stz2757}
M.~Le~Delliou, R.J.F. Marcondes, G.B. Lima~Neto, Monthly Notices of the Royal
  Astronomical Society \textbf{490}(2), 1944 (2019).
\newblock \doi{10.1093/mnras/stz2757}

\bibitem{Riess_1998}
A.G. Riess, others., The Astronomical Journal \textbf{116}(3), 1009–1038
  (1998).
\newblock \doi{10.1086/300499}

\bibitem{10.1093/mnras/150.1.1}
H.A. Buchdahl, Monthly Notices of the Royal Astronomical Society
  \textbf{150}(1), 1 (1970).
\newblock \doi{10.1093/mnras/150.1.1}

\bibitem{STAROBINSKY198099}
A.A. Starobinsky, Physics Letters B \textbf{91}(1), 99 (1980).
\newblock \doi{https://doi.org/10.1016/0370-2693(80)90670-X}

\bibitem{Hehl_1995}
F.W. Hehl, J.D. McCrea, E.W. Mielke, Y.~Ne’eman, Physics Reports
  \textbf{258}(1–2), 1–171 (1995).
\newblock \doi{10.1016/0370-1573(94)00111-f}

\bibitem{Baldazzi_2022}
A.~Baldazzi, O.~Melichev, R.~Percacci, Annals of Physics \textbf{438}, 168757
  (2022).
\newblock \doi{10.1016/j.aop.2022.168757}

\bibitem{Vitagliano_2011}
V.~Vitagliano, T.P. Sotiriou, S.~Liberati, Annals of Physics \textbf{326}(5),
  1259–1273 (2011).
\newblock \doi{10.1016/j.aop.2011.02.008}

\bibitem{Karahan_2012}
C.N. Karahan, A.~Altaş, D.A. Demir, General Relativity and Gravitation
  \textbf{45}(2), 319–343 (2012).
\newblock \doi{10.1007/s10714-012-1473-x}

\bibitem{SARDANASHVILY_2011}
G.~Sardanashvily, International Journal of Geometric Methods in Modern Physics
  \textbf{08}(08), 1869–1895 (2011).
\newblock \doi{10.1142/s0219887811005993}

\bibitem{OLMO_2011}
G.J. Olmo., International Journal of Modern Physics D \textbf{20}(04),
  413–462 (2011).
\newblock \doi{10.1142/s0218271811018925}

\bibitem{ASENS_1924_3_41__1_0}
E.~Cartan, Annales scientifiques de l'\'Ecole Normale Sup\'erieure \textbf{3e
  s{\'e}rie, 41}, 1 (1924).
\newblock \doi{10.24033/asens.753}

\bibitem{ASENS_1925_3_42__17_0}
E.~Cartan, Annales scientifiques de l'\'Ecole Normale Sup\'erieure \textbf{3e
  s{\'e}rie, 42}, 17 (1925).
\newblock \doi{10.24033/asens.761}

\bibitem{Kaluza_Klein}
O.~Klein, Zeitschrift f{\"u}r Physik \textbf{37}(12), 895 (1926).
\newblock \doi{10.1007/BF01397481}

\bibitem{Klein}
O.~Klein, Zeitschrift f{\"u}r Physik \textbf{37}(12), 895 (1926).
\newblock \doi{10.1007/BF01397481}

\bibitem{saridakis2023modified}
E.N. Saridakis, et~al.
\newblock Modified gravity and cosmology: An update by the {CANTATA} network
  (2023)

\bibitem{Shankaranarayanan_2022}
S.~Shankaranarayanan, J.P. Johnson, General Relativity and Gravitation
  \textbf{54}(5) (2022).
\newblock \doi{10.1007/s10714-022-02927-2}

\bibitem{Olmo:2011uz}
G.J. Olmo, Int. J. Mod. Phys. D \textbf{20}, 413 (2011).
\newblock \doi{10.1142/S0218271811018925}

\bibitem{Bahamonde:2015zma}
S.~Bahamonde, C.G. B\"ohmer, M.~Wright, Phys. Rev. D \textbf{92}(10), 104042
  (2015).
\newblock \doi{10.1103/PhysRevD.92.104042}

\bibitem{Bahamonde:2022kwg}
S.~Bahamonde, J.~Chevrier, J.~Gigante~Valcarcel, JCAP \textbf{02}, 018 (2023).
\newblock \doi{10.1088/1475-7516/2023/02/018}

\bibitem{Bahamonde:2021srr}
S.~Bahamonde, A.~Golovnev, M.J. Guzm\'an, J.L. Said, C.~Pfeifer, JCAP
  \textbf{01}(01), 037 (2022).
\newblock \doi{10.1088/1475-7516/2022/01/037}

\bibitem{BeltranJimenez:2017tkd}
J.~Beltr\'an~Jim\'enez, L.~Heisenberg, T.~Koivisto, Phys. Rev. D
  \textbf{98}(4), 044048 (2018).
\newblock \doi{10.1103/PhysRevD.98.044048}

\bibitem{Harko:2018gxr}
T.~Harko, T.S. Koivisto, F.S.N. Lobo, G.J. Olmo, D.~Rubiera-Garcia, Phys. Rev.
  D \textbf{98}(8), 084043 (2018).
\newblock \doi{10.1103/PhysRevD.98.084043}

\bibitem{Silva:2022pfd}
J.E.G. Silva, R.V. Maluf, G.J. Olmo, C.A.S. Almeida, Phys. Rev. D
  \textbf{106}(2), 024033 (2022).
\newblock \doi{10.1103/PhysRevD.106.024033}

\bibitem{Hehl:1994ue}
F.W. Hehl, J.D. McCrea, E.W. Mielke, Y.~Ne'eman, Phys. Rept. \textbf{258}, 1
  (1995).
\newblock \doi{10.1016/0370-1573(94)00111-F}

\bibitem{Eddington1923-EDDTMT}
A.S. Eddington, \emph{The Mathematical Theory of Relativity} (The University
  Press, Cambridge [Eng.], 1923)

\bibitem{schrodinger1985space}
E.~Schr{\"o}dinger, \emph{Space-Time Structure}.
\newblock Cambridge Science Classics (Cambridge University Press, 1985)

\bibitem{Einstein:1923:AFG}
A.~Einstein, Sitzungsber. Preuss. Akad. Wiss. pp. 137--140 (1923)

\bibitem{eisenhart1972non}
L.P. Eisenhart, \emph{Non-Riemannian Geometry} (American Mathematical Society,
  1972)

\bibitem{POP_AWSKI_2007}
N.J. Poplawski, Modern Physics Letters A \textbf{22}(36), 2701–2720 (2007).
\newblock \doi{10.1142/s0217732307025662}

\bibitem{Knorr_2021}
B.~Knorr, C.~Ripken, Physical Review D \textbf{103}(10) (2021).
\newblock \doi{10.1103/physrevd.103.105019}

\bibitem{POP_AWSKI_2008}
N.J. Poplawski, International Journal of Modern Physics A \textbf{23}(3-4),
  567–579 (2008).
\newblock \doi{10.1142/s0217751x08039578}

\bibitem{Pop_awski_2009}
N.J. Popławski, Foundations of Physics \textbf{39}(3), 307–330 (2009).
\newblock \doi{10.1007/s10701-009-9284-y}

\bibitem{Filippov_2010}
A.T. Filippov, Theoretical and Mathematical Physics \textbf{163}(3), 753–767
  (2010).
\newblock \doi{10.1007/s11232-010-0059-6}

\bibitem{Azri_2015}
H.~Azri, Classical and Quantum Gravity \textbf{32}(6), 065009 (2015).
\newblock \doi{10.1088/0264-9381/32/6/065009}

\bibitem{Krasnov_2011}
K.~Krasnov, Physical Review Letters \textbf{106}(25) (2011).
\newblock \doi{10.1103/physrevlett.106.251103}

\bibitem{BI_Gravity}
M.~Born, L.~Infeld, Nature \textbf{132}(3348), 1004 (1933).
\newblock \doi{10.1038/1321004b0}

\bibitem{Deser_1998}
S.~Deser, G.W. Gibbons, Classical and Quantum Gravity \textbf{15}(5), L35–L39
  (1998).
\newblock \doi{10.1088/0264-9381/15/5/001}

\bibitem{Vollick:2003qp}
D.N. Vollick, Phys. Rev. D \textbf{69}, 064030 (2004).
\newblock \doi{10.1103/PhysRevD.69.064030}

\bibitem{Banados:2010ix}
M.~Banados, P.G. Ferreira, Phys. Rev. Lett. \textbf{105}, 011101 (2010).
\newblock \doi{10.1103/PhysRevLett.105.011101}.
\newblock [Erratum: Phys.Rev.Lett. 113, 119901 (2014)]

\bibitem{Jim_nez_2021}
J.B. Jiménez, A.~Delhom, G.J. Olmo, E.~Orazi, Physics Letters B \textbf{820},
  136479 (2021).
\newblock \doi{10.1016/j.physletb.2021.136479}

\bibitem{Afonso:2021aho}
V.I. Afonso, C.~Bejarano, R.~Ferraro, G.J. Olmo, Phys. Rev. D \textbf{105}(8),
  084067 (2022).
\newblock \doi{10.1103/PhysRevD.105.084067}

\bibitem{castillofelisola2016polynomial}
O.~Castillo-Felisola, A.~Skirzewski.
\newblock A polynomial model of purely affine gravity (2016)

\bibitem{castillofelisola2016einsteins}
O.~Castillo-Felisola, A.~Skirzewski.
\newblock Einstein's gravity from a polynomial affine model (2016)

\bibitem{castillofelisola2019cosmological}
O.~Castillo-Felisola, J.~Perdiguero, O.~Orellana.
\newblock Cosmological solutions to polynomial affine gravity in the
  torsion-free sector (2019)

\bibitem{Castillo_Felisola_2018}
O.~Castillo-Felisola, in \emph{Gravity - Geoscience Applications, Industrial
  Technology and Quantum Aspect} ({InTech}, 2018).
\newblock \doi{10.5772/intechopen.70951}

\bibitem{Castillo_Felisola_2020}
O.~Castillo-Felisola, J.~Perdiguero, O.~Orellana, A.R. Zerwekh, Classical and
  Quantum Gravity \textbf{37}(7), 075013 (2020).
\newblock \doi{10.1088/1361-6382/ab58ef}

\bibitem{Castillo_Felisola_2022_EPJC}
O.~Castillo-Felisola, O.~Orellana, J.~Perdiguero, F.~Ram{\'{\i}}rez,
  A.~Skirzewski, A.R. Zerwekh, The European Physical Journal C \textbf{82}(1)
  (2022).
\newblock \doi{10.1140/epjc/s10052-021-09938-4}

\bibitem{Castillo_Felisola_2022_Universe}
O.~Castillo-Felisola, B.~Grez, O.~Orellana, J.~Perdiguero, F.~Ramirez,
  A.~Skirzewski, A.R. Zerwekh, Universe \textbf{8}(2), 68 (2022).
\newblock \doi{10.3390/universe8020068}

\bibitem{Olmo:2022ops}
G.J. Olmo, E.~Orazi, G.~Pradisi, JCAP \textbf{10}, 057 (2022).
\newblock \doi{10.1088/1475-7516/2022/10/057}

\bibitem{castillofelisola2023inflationary}
O.~Castillo-Felisola, B.~Grez, Jose, A.~Skirzewski.
\newblock Inflationary scenarios in an effective polynomial affine model of
  gravity (2023)

\bibitem{Castillo-Felisola17}
O.~Castillo-Felisola, in \emph{Gravity}, ed. by T.~Zouaghi (IntechOpen, Rijeka,
  2017), chap.~9.
\newblock \doi{10.5772/intechopen.70951}

\bibitem{KJ_Formalism}
J.~Kijowski, General Relativity and Gravitation \textbf{9}(10), 857 (1978).
\newblock \doi{10.1007/BF00759646}

\bibitem{Castillo-Felisola_2023}
O.~Castillo-Felisola, B.~Grez, O.~Orellana, J.~Perdiguero, A.~Skirzewski, A.R.
  Zerwekh, Classical and Quantum Gravity \textbf{40}(24), 249501 (2023).
\newblock \doi{10.1088/1361-6382/ad0356}

\bibitem{Pop_awski_2013}
N.~Pop{\l}awski, General Relativity and Gravitation \textbf{46}(1) (2013).
\newblock \doi{10.1007/s10714-013-1625-7}

\bibitem{sagemath}
{The Sage Developers}, \emph{{S}ageMath, the {S}age {M}athematics {S}oftware
  {S}ystem ({V}ersion 10.3)} (2024).
\newblock {\tt https://www.sagemath.org}

\bibitem{Gourgoulhon_2015}
E.~Gourgoulhon, M.~Bejger, M.~Mancini, Journal of Physics: Conference Series
  \textbf{600}, 012002 (2015).
\newblock \doi{10.1088/1742-6596/600/1/012002}

\bibitem{Gourgoulhon_2018}
{\'{E}}.~Gourgoulhon, M.~Mancini, Les cours du {CIRM} \textbf{6}(1), 1 (2018).
\newblock \doi{10.5802/ccirm.26}

\bibitem{peeters2018introducing}
K.~Peeters.
\newblock Introducing cadabra: a symbolic computer algebra system for field
  theory problems (2018)

\bibitem{Peeters2018}
K.~Peeters, Journal of Open Source Software \textbf{3}(32), 1118 (2018).
\newblock \doi{10.21105/joss.01118}

\bibitem{Peeters_2007}
K.~Peeters, Computer Physics Communications \textbf{176}(8), 550 (2007).
\newblock \doi{10.1016/j.cpc.2007.01.003}

\end{thebibliography}

\end{document}